# Control of single-spin magnetic anisotropy by exchange coupling

Jenny C. Oberg[1,2,*], M. Reyes Calvo[1,*,†], Fernando Delgado[3], María Moro-Lagares[4,5], David Serrate[4,5], David Jacob[6], Joaquín Fernández-Rossier[3] & Cyrus F. Hirjibehedin[1,2,7,*]

[1]London Centre for Nanotechnology, University College London (UCL), London WC1H 0AH, UK.

[2]Department of Physics & Astronomy, UCL, London WC1E 6BT, UK.

[3]International Iberian Nanotechnology Laboratory (INL), 4715-330 Braga, Portugal.

[4]Instituto de Nanociencia de Aragón (INA) and Laboratorio de Microscopías Avanzadas (LMA), Universidad de Zaragoza, 50018 Zaragoza, Spain.

[5]Departamento de Física de la Materia Condensada, Universidad de Zaragoza, 50009 Zaragoza, Spain.

[6]Max-Planck-Institut für Mikrostrukturphysik, 06120 Halle, Germany.

[7]Department of Chemistry, UCL, London WC1H 0AJ, UK.

[*]These authors contributed equally to this work.

[†]present address: Department of Physics, Stanford University, Stanford, CA 94305, USA.

**The properties of quantum systems interacting with their environment, commonly called open quantum systems, can be strongly affected by this interaction. While this can lead to unwanted consequences, such as causing decoherence in qubits used for quantum computation[1], it can also be exploited as a probe of the environment. For example, magnetic resonance imaging is based on the dependence of the spin relaxation times of protons[2] in water molecules in a host's tissue[3]. Here we show that the excitation energy of a single spin, which is determined by magnetocrystalline anisotropy and controls its stability and suitability for use in magnetic data storage devices[4], can be modified by varying the exchange coupling of the spin to a nearby conductive electrode. Using scanning tunnelling microscopy and spectroscopy, we observe variations up to a factor of two of the spin excitation energies of individual atoms as the strength of the spin's coupling to the surrounding electronic bath changes. These observations, combined with calculations, show that exchange coupling can strongly modify the magnetic anisotropy. This system is thus one of the few open quantum systems in which the energy levels, and not just the excited-state lifetimes, can be controllably**

**renormalized. Furthermore, we demonstrate that the magnetocrystalline anisotropy, a property normally determined by the local structure around a spin, can be electronically tuned. These effects may play a significant role in the development of spintronic devices[5] in which an individual magnetic atom or molecule is coupled to conducting leads.**

In quantum mechanical systems, whenever coupling to the environment induces changes in the lifetimes of states it must also induce a shift (often referred to as "dressing" or renormalization) of the energy levels of the system[6]. Measuring the shifts, as opposed to the lifetimes, is difficult because it is often not straightforward to extract the bare energy from the dressed value obtained from spectroscopic techniques. Furthermore, the effect of the environment can go far beyond the renormalization of the energy levels. This occurs for instance in Kondo systems[7], where a localized spin is exchange coupled to a bath of itinerant electrons, screening the localized spin through the formation of a total spin singlet state together with the itinerant electrons.

The structure of the environment also influences open quantum systems. One very important and technologically relevant example of this is magnetic anisotropy. The push to increase data storage capacities to the ultimate limit[8] has driven research into understanding magnetic anisotropy at the atomic scale[9-14]. Tuning magnetic anisotropy normally can be done via structural[10,13] or mechanical[15] means, though electrical control of anisotropy through the addition and subtraction of discrete units of charge on a molecule has been observed[16]. The interplay between magnetic anisotropy and Kondo screening at the atomic and molecular scale has also recently received theoretical and experimental attention[15,17-19].

In our experiments (Methods Summary), Co atoms are deposited on a thin-decoupling layer of copper nitride ($Cu_2N$) created on Cu(001). $Cu_2N$ reduces the coupling of magnetic atoms with the underlying metallic substrate[11,18]. As seen in the scanning tunnelling microscopy (STM) image shown in Fig. 1a, the $Cu_2N$ islands used here are significantly larger than those used in some prior experiments[11,18]. Scanning tunnelling spectroscopy (STS) measurements performed on four representative atoms on this island are shown in Fig. 1b; note that the atoms have negligible differences in their topographic appearance at the voltage at which they were imaged. In these spectra, two distinct features are seen: a peak in the local density of states (LDOS) centred at zero bias and two steps in differential conductance that occur symmetrically at positive and negative voltages. In prior experiments[18], the zero-bias peak was found to be a Kondo resonance while the differential conductance steps were inelastic electron tunnelling (IET) transitions at spin excitation energies described by the spin Hamiltonian[4]

$$H = g\mu_B \vec{B} \cdot \vec{S} + DS_z^2 + E\left(S_x^2 - S_y^2\right) \tag{1}$$

where $\mu_B$ is the Bohr magneton, $g$ is the Landé g-factor, $B$ is the magnetic field, $D$ and $E$ are the axial and transverse anisotropy constants, $S=3/2$ is the total spin, and $S_{x,y,z}$ are the projections of the spin along the appropriate axes.

The most striking result of this work is that, as observed in Fig. 1b, the spectra of the different Co atoms on the $Cu_2N$ change dramatically even though the atoms are simply at different positions on the same surface, with no observed changes in the local binding. At the edge of a large (18.6×20.5 nm$^2$) $Cu_2N$ island, the STS spectrum of Co closely resembles prior measurements for Co on small (5×5 nm$^2$) $Cu_2N$ islands[18]. However, as the atom's position shifts towards the centre, two striking changes occur: the relative height of the Kondo peak decreases, and the IET step shifts to significantly higher energy. Because the IET step is a measure of the magnetic anisotropy energy, this suggests that the anisotropy energy is increasing as the Kondo screening is decreasing.

A first candidate to account for the observed variations in the magnetocrystalline anisotropy would be a change in the structure of the $Cu_2N$. Both magnetic anisotropy and exchange coupling arise from the overlap of the orbitals of the Co atom with those of the atoms in the surface, primarily the neighbouring nitrogen atoms[11]. For small crystal deformations, the relative change of the magnetocrystalline anisotropy should be proportional to the strain, with the constant of proportionality of order unity in the elastic regime[4,20]. Atomically resolved images of the $Cu_2N$ islands reveal no change in the lattice constant (with an uncertainty of a few pm), limiting the maximum strain to be a few percent. Measurements of the Cu$_2$N bandgap onset, whose change should be very sensitive to small changes in the lattice constant[21], further restrict the maximum size of the strain: as seen in Fig. 1c, we find that the bandgap shifts by a few percent at the centre of the island compared to the edges, suggesting substantially smaller changes in the lattice. Therefore, while structural changes of the Cu$_2$N may account for some portion (less than 1%) of our observed energy shift, they cannot account for the majority of the effect. Detailed and technically demanding calculations of large Cu$_2$N islands may provide a more precise measure of this contribution.

Having ruled out a structural origin of the joint variations of Kondo resonance and the effective magnetic anisotropy, we explore a new physical scenario where changes in the exchange coupling between the Co spin and the conduction electrons, which lead naturally to a change of the Kondo temperature, also affect the spin excitation energies. We do so using both the Kondo and Anderson models, generalized to include single ion magnetic anisotropy (Supplementary Discussion). In the Kondo model, the dimensionless constant $\rho J$, the product of the density of states of substrate electrons at the Fermi energy $\rho$ and the exchange energy $J$, controls the influence of the conduction electrons: the spin susceptibility is renormalized to linear order in $\rho J$ while the local spin relaxation rate is proportional to $(\rho J)^2$ [22-24]. In nuclear magnetic resonance, these phenomena are the well-known Knight shift and Korringa spin relaxation, respectively[25]. The environmentally induced decay rate necessarily comes together with a renormalization of the associated transition energy[6]. For the Kondo model with single ion anisotropy, second order perturbation theory yields the following expression for the renormalized excitation energy (Supplementary Discussion):

$$\Delta = \Delta_0 \left(1 - \tfrac{3}{16}(\rho J)^2 \ln \frac{2W}{\pi k_B T}\right) \qquad (2)$$

where $\Delta_0$ is the bare excitation energy, corresponding to spin excitations between the levels described by equation (1); $k_B$ is Boltzmann's constant; $T$ is the temperature; and $W$ is the bandwidth of the substrate electrons. The second term in this equation is an exchange-driven shift of the spin excitation energies and is formally similar to the normally overlooked second order contribution to the Knight shift[26]. Qualitatively equation (2) accounts for our central observation: as $\rho J$ decreases, the Kondo temperature[7] $T_K$

$$k_B T_K \propto W\sqrt{\rho J} e^{-1/\rho J} \qquad (3)$$

goes down while at the same time the spin excitation energy goes up. Whereas in most systems environmentally induced shifts can not be quantified because it is not possible to determine the bare energy $\Delta_0$, here the correlated variations of the Kondo temperature and the excitation energy reveal the significant renormalization of the single ion magnetic anisotropy by Kondo exchange.

Equation (2) is based on a perturbative calculation and, as such, cannot reproduce the full Kondo phenomenology. To overcome this limitation, obtain further evidence for the above scenario, and have a more microscopic understanding of the origin of the variation of $\rho J$, we have carried out non-perturbative calculations, based upon a multi-orbital Anderson model with three local orbitals holding an anisotropic spin-3/2 (Supplementary Discussion). This model is defined by three parameters: the one-electron energies of the local orbitals $E_d$, the effective Coulomb repulsion between electrons $U$, and the single-particle broadening $\Gamma$ due to tunnelling between the local orbitals and the substrate. $E_d$ and $U$ determine the electron removal and addition energies $E_0$ and $U-E_0$ (Supplementary Discussion); therefore $E_0$, $U-E_0$, and $\Gamma$ are the relevant energy scales that govern the Kondo physics.

We solve the generalized Anderson model with the One Crossing Approximation (OCA)[27] and obtain the spectral function, which can be related to the experimentally measured STS spectra[28]. The observed symmetry upon bias inversion of the experimental $dI/dV$ curves is best reproduced when we consider the electron-hole symmetric case ($E_0=U/2$), as illustrated in Fig. 2; however, our results are robust and also occur in the absence of electron-hole symmetry. In the symmetric limit, the relation between the Anderson and Kondo models leads to a particularly simple linear relationship[7]: $\rho J = 8\,\Gamma/U$. Thus, a change in $\rho J$ can arise in general from variations of $\Gamma$, $U$, or $E_d$.

Figures 1d and 2 highlight the results of our OCA calculations. Increasing $\Gamma$, keeping $U$ constant, the charge addition peaks at high energy broaden and shift (Fig. 2c). Moreover, in an energy window of ~10 meV around the Fermi energy, two relevant features are found, in agreement with our experimental observations: a Kondo resonance at the Fermi energy and a step a few meV above and below. Our OCA calculations show that the Kondo peak grows as $\Gamma$ increases, while at the same time the spin excitation step shifts to lower energy (Supplementary Discussion), in agreement with the perturbative theory. As illustrated in Fig. 2d, this general behaviour in our OCA calculations is not sensitive to the specific choice of $D$ or $E$. Importantly, the non-perturbative results show that the shift also changes linearly with $(\Gamma/U)^2$, in qualitative agreement with equation (2). A similar shift of the singlet to triplet excitation energy has been recently obtained from an Anderson model of two exchange coupled spin ½ sites treated in the Non Crossing Approximation[29].

Renormalization of the magnetic anisotropy can arise in a variety of different scenarios where $\Gamma$, $U$, and $E_d$ change at different locations on the surface. Here, we believe that variations in $\Gamma$ are the most likely cause of the observed changes of the $dI/dV$. As seen in Fig. 1c, the gap of $Cu_2N$ increases by about 0.1 V as we move from the island edge to the island centre. A larger gap implies a higher tunnelling barrier, leading to a smaller $\Gamma$ and therefore a smaller $\rho J$. However, a comparison with results obtained on islands with different sizes, showing that large islands present a variation of the magnetic anisotropy far from the edges despite the apparent constant gap, suggests that the situation may be more complex (Supplementary Discussion). For example, surface states confined under the $Cu_2N$ may play a role[30]. In addition, variations in $U$ and $E_d$, which have been correlated with substantial changes in Kondo screening for Co on Cu(100)[31], may also drive variations in exchange coupling. However, our calculations suggest that these parameters must change by more than 1 eV to account for a significant fraction of the observed shifts.

The magnetic field behaviour of the STS also changes as the position of a Co atom varies on the large $Cu_2N$ island. As seen in Fig. 3a,b,d,e, the field dependence of the IET step for a Co atom near the edge of the island is well-described by equation (1) with a large *D* term and *E*~0, consistent with results obtained at the centre of small $Cu_2N$ islands[18]. However, as seen in Fig. 3c,f, the IET step of a Co atom near the centre of the large $Cu_2N$ island can only be properly described by including a large *E* term. Excellent qualitative agreement between the spectral functions calculated using the OCA for the Anderson model (Fig. 1d) and the experimental spectra (Fig. 1b) are obtained using the values of *D* and *E* obtained in Fig. 3f.

We note that the Co atom's environment becomes more isotropic as $\Gamma$ increases. More precisely, for systems with both axial and transverse anisotropy, all three axes are different[4]. As $\Gamma$ increases, exchange will dominate the smaller transverse term, leaving the system with just a smaller axial anisotropy; eventually, for large $\Gamma$ the system will effectively become isotropic. The first stage of this is precisely what is observed experimentally in Fig. 3.

Exchange driven renormalization of magnetic anisotropy should be present in any system in which a magnetic impurity is coupled to an electronic bath, even if no Kondo screening occurs, but normally cannot be observed directly because either the unscreened spin excitations cannot be determined directly or the coupling cannot be controllably varied. Understanding this phenomenon is therefore crucial for future engineering of nanoscale quantum spintronic systems, which often involves placing an atomic or molecular spin in contact with an electronic reservoir[5]. Magnetic atoms on large $Cu_2N$ islands are therefore a special physical system with which we can observe and thereby understand the quantum mechanical "dressing" and "undressing" of a spin. This renormalization also provides an electronically tunable mechanism for controlling the magnetic anisotropy experienced by a quantum spin, which could have significant ramifications for the design and control of magnetic bits at the atomic and molecular scale. Not only does this mechanism enable control of the magnitude of the magnetic anisotropy, but it also can be used to tune the relative strengths of the axial and transverse terms, which can be used to enhance or weaken various charge and spin tunnelling phenomena[4,19].

**Methods Summary**

The majority of the STM experiments were performed using an Omicron Cryogenic STM operating in ultrahigh-vacuum (chamber pressures below $5\times10^{-10}$ mbar) at an effective sample temperature of 2.5 K. Superconducting magnets can apply fields of up to 6 T perpendicular to the surface of the sample or up to 2 T perpendicular to the surface of the sample plus up to 1 T in the plane. Additional STM experiments were performed using a SPECS JT-STM, a commercial adaptation of the design described by L. Zhang *et al.*[32], operating in ultrahigh-vacuum with similar chamber pressures and at a base temperature of 5 K.

Cu(001) samples (MaTeck single crystal with 99.999% purity) were prepared by repeated cycles of sputtering and annealing with Ar and annealing to 500°C. $Cu_2N$ was prepared on top of clean Cu(001) samples by sputtering with $N_2$ and annealing to 350°C. The sample was held below 30 K while Co atoms were evaporated onto the surface.

The bias voltage $V$ is always quoted in sample bias convention. Topographic images were obtained in the constant current imaging mode with $V$ and tunnel current $I$ set to $V_0$ and $I_0$ respectively and processed using WSxM[33]. Differential conductance measurements were obtained using a lock-in amplifier, with AC modulation voltages of 100 μV at approximately 750 Hz added to $V$; spectra were acquired by initially setting $V = V_0$ and $I = I_0$, holding the tip at a fixed position above the surface, and then sweeping $V$ while recording $I$ and $dI/dV$.

Differential conductance spectra shown in Figs 1b-c and 3a-c and Supplementary Figs 3b-c and 4d taken at zero perpendicular field $B_\perp$ were acquired with an in-plane 1 T field to reduce vibrational noise; no noticeable change in the spectral features was observed compared to $B=0$.

**Supplementary Information** is linked to the online version of the paper at www.nature.com/nature.

**Acknowledgments** We acknowledge Benjamin E.M. Bryant, Andrew J. Fisher, Katharina J. Franke, Andreas J. Heinrich, Mark Hybertsen, Sebastian Loth, and Alexander F. Otte for discussions and Benjamin E.M. Bryant for support during the experiments. J.F.R. acknowledges the hospitality of the Max-Planck-Institut für Mikrostrukturphysik Halle. Also, J.F.R. is on leave from Departamento de Física Aplicada, Universidad de Alicante, Spain. This work was supported by the EPSRC (EP/D063604/1 and EP/H002022/1); MEC-Spain (FIS2010-21883-C02-01, MAT2010-19236, CONSOLIDER CSD2007-0010, and Programa de Movilidad Postdoctoral); European Commission FP7 programme (PER-GA-2009-251791); and GV grant Prometeo 2012-11.

**Author Contributions** J.C.O, M.R.C., and C.F.H. conceived of the experiments. J.C.O. and M.R.C. performed the primary experiments and the data analysis supervised by C.F.H. Additional experiments were performed by J.C.O with the assistance of M.M. and supervised by D.S. and C.F.H. F.D. performed the perturbative calculations of exchange-induced modification of magnetic anisotropy, as proposed by J.F.R. D.J. implemented and performed the Anderson model calculations in the one-crossing approximation as proposed by J.F.R. All authors discussed the results and participated in writing the manuscript.

**Author Information** Reprints and permissions information is available at www.nature.com/reprints. The authors declare no competing financial interests. Correspondence and requests for materials should be addressed to C.F.H. (c.hirjibehedin@ucl.ac.uk).

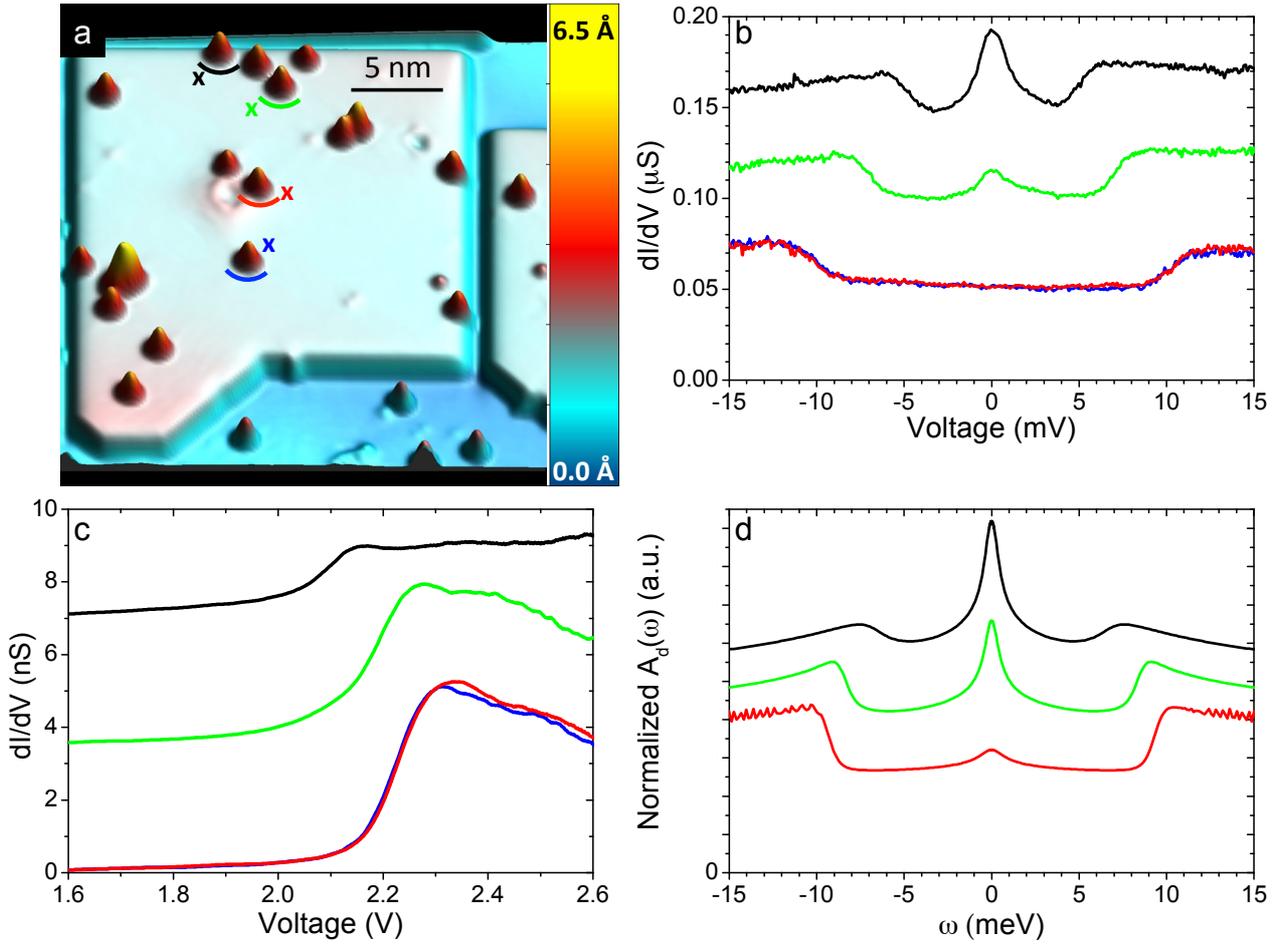

**Figure 1 | Spectroscopy of Co on a large (18.6×20.5 nm$^2$) Cu$_2$N island. a,** Topographic STM image (setpoint voltage $V_0$=100 mV, setpoint current $I_0$=100 pA) of Co atoms on a Cu$_2$N island. Coloured arcs label the atoms and crosses indicate the location of spectra acquired over nearby bare Cu$_2$N. **b,** Low-bias $dI/dV$ ($V_0$=15 mV, $I_0$=1 nA) spectroscopy acquired at perpendicular magnetic field $B_\perp$=0 on top of four atoms labelled in panel a; spectra are offset vertically for clarity. **c,** High-bias differential conductance spectra ($V_0$=1.5 V, $I_0$=50 pA) acquired at $B_\perp$=0 near atoms at locations labelled with a cross in panel a; spectra are offset vertically for clarity. **d,** Spectral function $A_d(\omega)$ obtained from the Anderson model calculations ($D$=3.5 meV, $E$=2 meV, $T$~2.3 K, $U$=4 eV) with $\Gamma$=20 meV (red), 50 meV (green), and 90 meV (black). For consistency with the STM spectra in panel b, $A_d(\omega)$ is normalized such that the integrated weight up to 15 mV is constant; spectra are vertically offset for clarity.

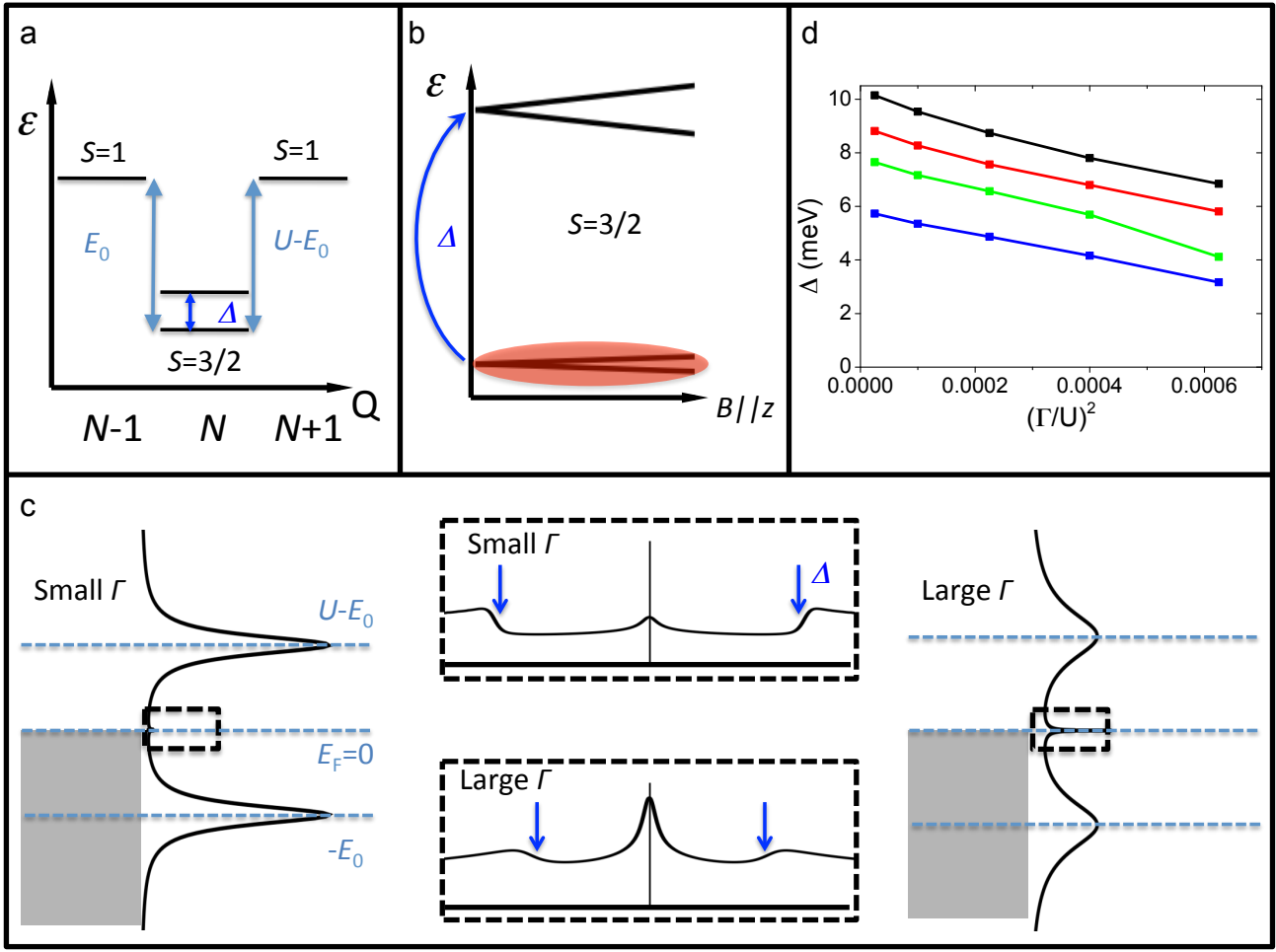

**Figure 2 | Generalized Anderson model of the Co electrons coupled to a bath of conduction electrons. a,** Scheme of the many body energy levels for the three charge states of Co. **b,** Scheme of the $S=3/2$ multiplet split by the magnetic anisotropy. The lowest (shaded) states form an effective two-level Kondo system. **c,** Scheme of the generalized Anderson model for small $\Gamma$ (left panel) and large $\Gamma$ (right panel), showing the addition energies and the spectral function $A_d(\omega)$; the gray shaded area represent the Fermi sea of conduction electrons. Fine structure around the Fermi energy $E_F$ is shown in the middle sections, with the blue vertical arrows labelling steps in the spectral function corresponding to spin excitations. **d,** Spin excitation energies obtained from the OCA for various anisotropy values (blue: $D=3$ meV, $E=0$ meV; red: 3 meV, 2 meV; green: 4 meV, 0 meV; black: 4 meV, 2 meV) and values of $(\Gamma/U)^2$.

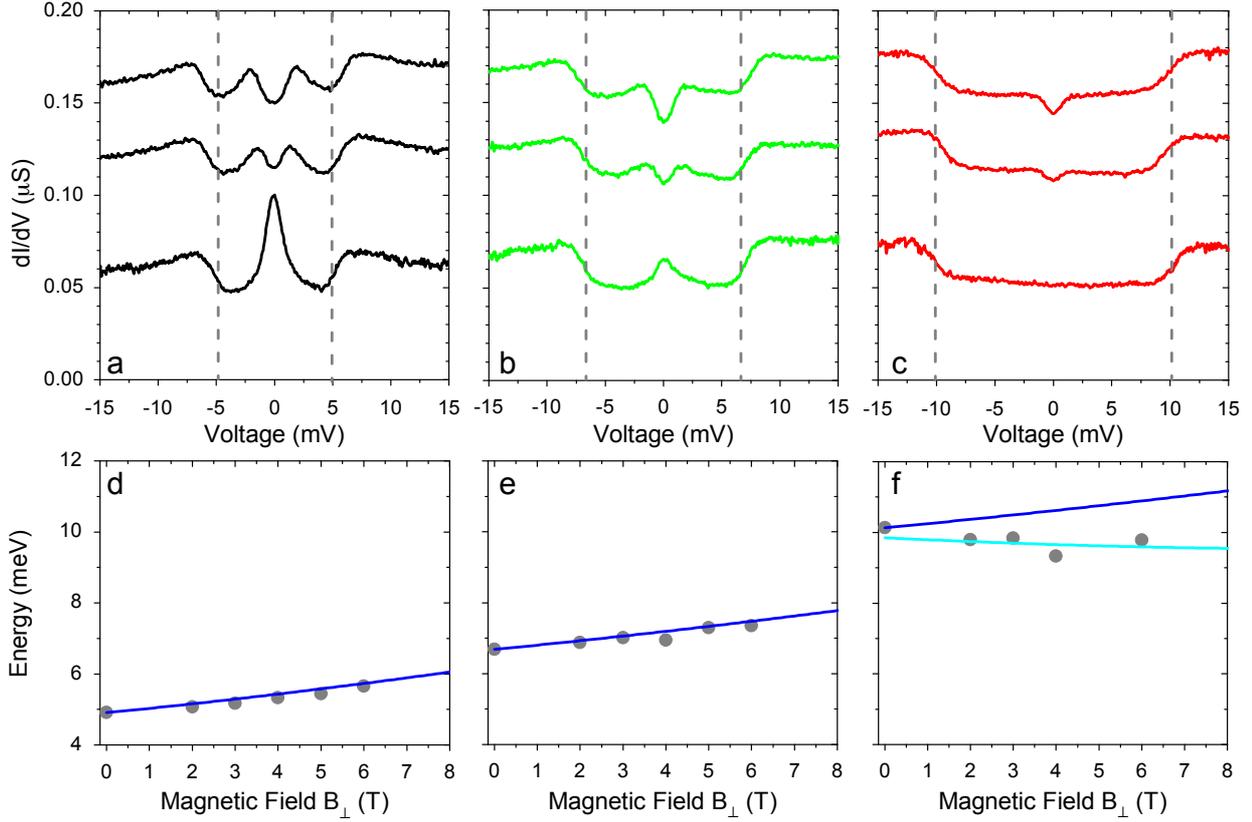

**Figure 3 | Magnetic field dependence of differential conductance spectra of Co on 18.6×20.5 nm² Cu₂N island. a-c,** Low bias differential conductance (*dI/dV*) spectra acquired at $B_\perp$=6 T (top), 4 T (middle), and 0 T (bottom) over atoms corresponding to those with similar colour labels in Fig. 1 ($V_0$=15 mV, $I_0$=1 nA); spectra are offset vertically for clarity and dashed vertical lines are a guide to the eye highlighting the change in energy of the IET step. **d-f,** IET step energy vs. perpendicular magnetic field. Solid dark blue line illustrates the evolution of equation (1) with *S*=3/2, *g*=2, *E*=0, and *D*=2.5 meV, 3.3 meV, and 5.0 meV (assigned based on the excitation energy at $B_\perp$=0) respectively; solid light blue line is for *D*=3.5 meV and *E*=2.0 meV, obtained from a fit of all the data points in panel f.